\documentclass[10pt,twoside,journal]{IEEEtran}
\usepackage[babel,german=quotes]{csquotes}

\usepackage{graphicx}  
\usepackage[caption=false]{subfig}
\usepackage{stfloats}
\usepackage{textpos}

\usepackage{gensymb}
\usepackage{amsmath}
\usepackage{tabularx}
\usepackage{footnote}
\usepackage[symbol]{footmisc}
\usepackage{hyperref}
\usepackage{wrapfig}
\usepackage{tcolorbox}
\usepackage{lipsum}

\hypersetup{
    colorlinks,
    linkcolor={black},
    citecolor={blue},
    urlcolor={blue}
   }
\usepackage{xcolor}
\usepackage{tikz}
\usetikzlibrary{arrows.meta}
\usetikzlibrary{%
    decorations.pathreplacing,%
    decorations.pathmorphing%
}
\usepackage{verbatim}
\usepackage[american,cuteinductors,smartlabels]{circuitikz}
\usetikzlibrary{math}
\usepackage{amsmath}
\tikzset{%
    body/.style={inner sep=0pt,outer sep=0pt,shape=rectangle,draw,thick,pattern=north east lines wide},
    dimen/.style={<->,>=latex,thin,every rectangle node/.style={fill=white,midway,font=\sffamily}},
    symmetry/.style={dashed,thin},
}
   
\def\TL#1#2#3#4#5{
\begin{scope}[shift={#1}, rotate=#2,scale=#3]
	\draw (0,0)--(1,0);
	\draw [fill={#4}] (1,-0.5) rectangle (5,0.5) node[pos=0.5,rotate=#2] {#5};
    \draw (5,0)--(6,0);
\end{scope}}

\def\openTL#1#2#3#4#5{
\begin{scope}[shift={#1}, rotate=#2,scale=#3]
	\draw (0,0)--(1,0);
	\draw [fill={#4}] (1,-0.5) rectangle (4,0.5) node[pos=0.5,rotate=#2+180] {#5};

\end{scope}}

\def\openStub#1#2#3#4#5{
\begin{scope}[shift={#1}, rotate=#2,scale=#3]
	\draw (0,0)--(1,0);
	\draw [fill={#4}] (1,-0.5) rectangle (4,0.5) node[pos=0.5,rotate=#2] {#5};

\end{scope}}

\def\port#1#2#3#4#5{
\begin{scope}[shift={#1}, rotate=#2,scale=#3]
	\draw (0,0)--(1,0);
	\draw (1,0)--(1+#5,0);
	\draw  (1+#5+.5,0) circle (0.5) node{#4};
\end{scope}}

\def\coupledTLr#1#2#3#4#5{
\begin{scope}[shift={#1}, rotate=#2,scale=#3]
	\draw (0,0)--(1,0);
	\draw [fill={#4}] (1,-0.3) rectangle (5,0.3) node[above left,rotate=#2] {#5};
    \draw (5,0)--(5.5,0);
    \draw (5.5,0)--(5.5,-0.9);
    \draw (5.5,-0.9)--(5,-0.9);
    \draw [fill={#4}] (1,-1.2) rectangle (5,-0.6);
    \draw (0,-0.9)--(1,-0.9);
\end{scope}}
\def\coupledTLl#1#2#3#4#5{
\begin{scope}[shift={#1}, rotate=#2,scale=#3]
	\draw (0,0)--(-1,0);
	\draw [fill={#4}] (-1,-0.3) rectangle (-5,0.3) node[above right,rotate=#2] {#5};
    \draw (-5,0)--(-5.5,0);
    \draw (-5.5,0)--(-5.5,-0.9);
    \draw (-5.5,-0.9)--(-5,-0.9);
    \draw [fill={#4}] (-1,-1.2) rectangle (-5,-0.6);
    \draw (0,-0.9)--(-1,-0.9);
\end{scope}}

\begin{document}
\textblockcolour{blue!15}
\title{Design Guidelines and Applications for Dual-Band Rat-Race Couplers and Gysel Power Dividers with Unequal Amplitude Imbalances}
\author{Rakesh~Sinha\textsuperscript{\href{https://orcid.org/0000-0003-0592-8505}{\includegraphics[scale=0.1]{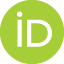}}},~\IEEEmembership{Member,~IEEE}
\vspace*{-1.2em}
\thanks{\color{red}This work has been accepted by the IEEE Microwave Magazine (\url{https://doi.org/10.1109/MMM.2024.3415241}) for possible
publication. Copyright may be transferred without notice, after which this
version may no longer be accessible. \color{black} This
work was supported by SERB under Grant SRG/2022/000156. (Corresponding
author: Rakesh Sinha.)}
\thanks{This paper has supplementary downloadable material available at
http://ieeexplore.ieee.org, provided by the author. This includes a PDF docu-
ment, containing additional design results generated using the free software available at \url{https://zenodo.org/records/11199141}. This material is 1.5
MB in size}
\thanks{Rakesh Sinha is with the Department of Electrical Engineering, National Institute of Technology, Rourkela-769008,
India (e-mail: r.sinha30@gmail.com, sinhar@nitrkl.ac.in ).
}

\thanks{Color versions of one or more of the figures in this paper are available online}}
\markboth{ACCEPTED AND TO APPEAR IN THE IEEE MICROWAVE MAGAZINE, }{SINHA: Design Guidelines and Applications for Dual-Band Rat-Race Couplers and Gysel Power Dividers with Unequal Amplitude Imbalances}
\maketitle
\begin{abstract}
This paper provides the design guidelines for a dual-band rat-race coupler and Gysel power divider with two different amplitude imbalances. The dual-band conditions are established based on the electrical nature of the coupler and dual-band C-section coupled lines/ T-structure/ $\Pi$-structures instead of using the dual-band equivalent transmission lines. It has been shown here that the same design equations can be utilized in the design of the rat-race coupler and Gysel power divider. The design equations are solved numerically to obtain the electrical parameters of the proposed coupler and power divider. The possible application of dual-band different power division ratios is presented for dual-band polarization controlled by a patch antenna. A design tool has been developed to obtain the coupler and power divider's design parameters and circuit analysis results. A design example of the coupler has been provided to validate the proposed concept.
\end{abstract}

\vspace*{-1em}
\section{Introduction}\label{sec1}
{Couplers and power dividers are integral parts of many microwave systems and sub-systems. Sometimes, these components are referred to as signal-separating structures because they separate the input signals into two output signals with different power labels and phase delays. In-phase/ out-of-phase couplers (rat-race couplers) and Gysel power dividers are two popular components because of their wide bandwidth and unequal power division flexibility. Here, we will show that the same design equations can be utilized to design both components. Modern-day communication requires multi-band components to harness different bands of the spectrum. Dual-band couplers with different power division ratios can find applications in the polarization control of dual-band antenna.}      

Couplers and power dividers are important fundamental passive components in the RF and microwave frequencies \cite{Sinha2016}. The design flexibility of equal or unequal power division is a crucial criterion for the couplers as different applications demand different power divisions; for example, the Nolen matrix \cite{Ren_Nolen_matrix_2019} requires couples with different power divisions. The dual-band or multi-band components required for modern-day communication and radar systems \cite{KMCheng2004}. In general, dual-band designs are achieved by replacing conventional transmission lines (TLs) with equivalent symmetric dual-band TLs \cite{Chi2009}. However, dual-band designs can also be achieved by exploiting the electrical nature of the coupler itself \cite{Yeung2011}-\cite{Wang2012} or by using the electrical nature of the coupler and dual-band TL theory \cite{Lwu2012}-\cite{Maktoomi2016}. In this context, electrical nature indicates that electrical lengths are proportional to frequency, and characteristic impedances are frequency invariant. Some of the commonly used equivalent dual-band TLs are $\Pi$-network  \cite{KMCheng2004}, \cite{Sinha_Cross_2016}, T-network \cite{Maktoomi2016}, \cite{Sinha_Cross_2016}, stepped impedance \cite{Lwu2012}, stepped-$\Pi$ \cite{Hsu2009}, T-network with stepped stub \cite{Chin2010}, parallel or dual-TL \cite{Gai2016}-\cite{Park2017}, and C-section coupled line (CCL) \cite{Chiou2009}-\cite{Tang2016}. 
 
 \color{black}Most dual-band couplers are designed for identical power division ratio at the two bands. However, some applications, like different polarization control at two frequencies, require non-identical power division ratios at two bands. To the best of this author's knowledge, only \cite{Lwu2012}, \cite{Chi2014} and \cite{ChenGPD2019} provide the solution to such problems. However, the double-layer strip technology used in \cite{Lwu2012} to implement the $180^{\circ}$ phase inverter is costly compared to single-layer microstrip technology. In \cite{Chi2014}, a dual-band non-quadrature branch-line coupler (BLC) was designed using dual-band-$\Pi$ structures. The additional stubs of the $\Pi$ structures are difficult to place inside the BLC, thus it requires a larger footprint. An alternative procedure is provided in \cite{ChenGPD2019}, where the electrical property of the Gysel power divider (GPD) and a dual-band double-$\Pi$ (for the phase inverter) are utilized. However, the double-$\Pi$ structure also requires an additional footprint area. 
 
This article provides design guidelines for the dual-band rat-race coupler (RRC) and GPD \cite{Tang2016}-\cite{Ahn2019} with different power division ratios at the design frequencies. {The designs are based on the general theory of RRC \cite{Sinha2016} and dual-band C-section coupled line (CCL) \cite{Tang2016} or Schiffman phase-shifter \cite{Schiffman_Phase_shifter58} as $90^{\circ}$ phase shifters. In addition, it is possible to use T-structures \cite{Maktoomi2016}, \cite{Sinha_Cross_2016} and $\Pi$-structures \cite{KMCheng2004}, \cite{Sinha_Cross_2016} as $90^{\circ}$ phase shifters. Based on the proposed design guidelines, the prototype can be easily realized using single-layer microstrip technology. The computed results for the RRC case are verified experimentally.}

{A computer-aided design application was developed to ease the computation of the design parameters and performance measures of the proposed coupler and power divider. The application is freely available at \url{https://zenodo.org/records/11199141} under Creative Commons License 4.0. Interested readers are encouraged to explore the applications.}    

The rest of the manuscript is organized as follows. Dual-band RRC is discussed in Section-\ref{sec2}. In Section-\ref{sec3}, the dual-band GPD is designed using the RRC design equations. The application of the resulting dual-band nonidentical power division device is illustrated in Section-\ref{sec4}. A brief description of the free software developed for this article is provided in Section-\ref{sec:cad}. In Section-\ref{sec5}, an example of RRC is provided.  Finally, the article concludes in Section-\ref{sec6}.  \color{black}

\begin{figure}
\centering
\subfloat[]{
\begin{tikzpicture}
  \node[rotate=0,scale=0.8] at (0,0) {
\begin{tikzpicture}[scale=0.8]
\tikzmath{\x1 = 3;\s=0.5;} 
\TL{(0,0)}{0}{\s}{red!50}{$Z_{\alpha},\; \theta_{\alpha}$};
\TL{(0,0)}{-90}{\s}{blue!50}{$Z_{\beta},\; \theta_{\beta}$};
\TL{(\x1,0)}{-90}{\s}{blue!50}{$Z_{\beta},\; \theta_{\beta}$};
\port{(0,0)}{180}{\s}{1}{0.5};
\port{(\x1,0)}{0}{\s}{3}{0.5};
\port{(0,-\x1)}{180}{\s}{2}{0.5};
\port{(\x1,-\x1)}{0}{\s}{4}{0.5};
\coupledTLr{(0,-1*\x1)}{45}{0.3}{orange!80}{};
\coupledTLl{(\x1,-1*\x1)}{-45}{0.3}{orange!80}{};
\TL{(0.2,-1*\x1-0.2)}{0}{0.435}{red!50}{$Z_{\alpha},\;  \theta_{\alpha}$};
\draw[] (1.5,-1.5) node[] {$Z_{\delta e}, Z_{\delta o},\; \theta_{\delta}$};
\end{tikzpicture}\label{fig:dual_rrc_c}
};
\end{tikzpicture}
}
\subfloat[]{
\begin{tikzpicture}
  \node[rotate=90,scale=0.8] at (0,0) {
\begin{tikzpicture}[scale=0.8]
\tikzmath{\x1 = 3;\s=0.5;} 
\TL{(0,0)}{0}{\s}{red!50}{$Z_{\alpha},\; \theta_{\alpha}$};
\TL{(0,0)}{-90}{\s}{blue!50}{$Z_{\beta},\; \theta_{\beta}$};
\TL{(\x1,0)}{-90}{\s}{blue!50}{$Z_{\beta},\; \theta_{\beta}$};
\port{(0,0)}{180}{\s}{1}{0.5};
\port{(\x1,0)}{0}{\s}{3}{0.5};
\port{(0,-\x1)}{180}{\s}{2}{0.5};
\port{(\x1,-\x1)}{0}{\s}{4}{0.5};
\draw (1,-1*\x1+0.5) to [short] (0,-1*\x1); 
\draw (\x1-1,-1*\x1+0.5) to [short] (\x1,-1*\x1); 
\openTL{(1,-1*\x1+0.5)}{90}{0.5}{green!80}{$Z_{2\pi},\; \theta_{2\pi}$};
\openTL{(\x1-1,-1*\x1+0.5)}{90}{0.5}{green!80}{$Z_{2\pi},\; \theta_{2\pi}$};
\TL{(0,-2*\x1)}{0}{\s}{red!50}{$Z_{\alpha},\;  \theta_{\alpha}$};
\TL{(0,-\x1)}{-90}{\s}{pink!80}{$Z_{1\pi},\; \theta_{1\pi}$};
\TL{(\x1,-\x1)}{-90}{\s}{pink!80}{$Z_{1\pi},\; \theta_{1\pi}$};
\draw (\x1-1,-2*\x1+1) to [short] (\x1,-2*\x1);
\draw (1,-2*\x1+1) to [short] (0,-2*\x1);
\openTL{(\x1-1,-2*\x1+1)}{90}{\s}{green!80}{$Z_{2\pi},\; \theta_{2\pi}$};
\openTL{(1,-2*\x1+1)}{90}{\s}{green!80}{$Z_{2\pi},\; \theta_{2\pi}$};
\end{tikzpicture}
\label{fig:dual_rrc_pi}};
\end{tikzpicture}
}\\
\subfloat[]{
\begin{tikzpicture}
  \node[rotate=90,scale=0.8] at (0,0) {
\begin{tikzpicture}[scale=0.8]
\tikzmath{\x1 = 3;\s=0.5;} 
\TL{(0,0)}{0}{\s}{red!50}{$Z_{\alpha},\; \theta_{\alpha}$};
\TL{(0,0)}{-90}{\s}{blue!50}{$Z_{\beta},\; \theta_{\beta}$};
\TL{(\x1,0)}{-90}{\s}{blue!50}{$Z_{\beta},\; \theta_{\beta}$};
\port{(0,0)}{180}{\s}{1}{0.5};
\port{(\x1,0)}{0}{\s}{3}{0.5};
\port{(0,-\x1)}{180}{\s}{2}{0.5};
\port{(\x1,-\x1)}{0}{\s}{4}{0.5};
\TL{(0,-3*\x1)}{0}{\s}{red!50}{$Z_{\alpha},\;  \theta_{\alpha}$};
\TL{(0,-\x1)}{-90}{\s}{pink!80}{$Z_{1t},\; \theta_{1t}$};
\TL{(\x1,-\x1)}{-90}{\s}{pink!80}{$Z_{1t},\; \theta_{1t}$};
\TL{(0,-2*\x1)}{-90}{\s}{pink!80}{$Z_{1t},\; \theta_{1t}$};
\TL{(\x1,-2*\x1)}{-90}{\s}{pink!80}{$Z_{1t},\; \theta_{1t}$};
\openTL{(\x1,-2*\x1)}{150}{\s}{cyan!80}{$Z_{2t},\; \theta_{2t}$};
\openTL{(0,-2*\x1)}{-30}{\s}{cyan!80}{$Z_{2t},\; \theta_{2t}$};
\end{tikzpicture}
\label{fig:dual_rrc_t}};
\end{tikzpicture}
}
\caption{Dual-band RRCs using TLs and dual-band quadrature lines: (a) C-section coupled line, (b) Pi-structure, and (c) T-structure. }
\label{fig:dual_rrc}
\end{figure}
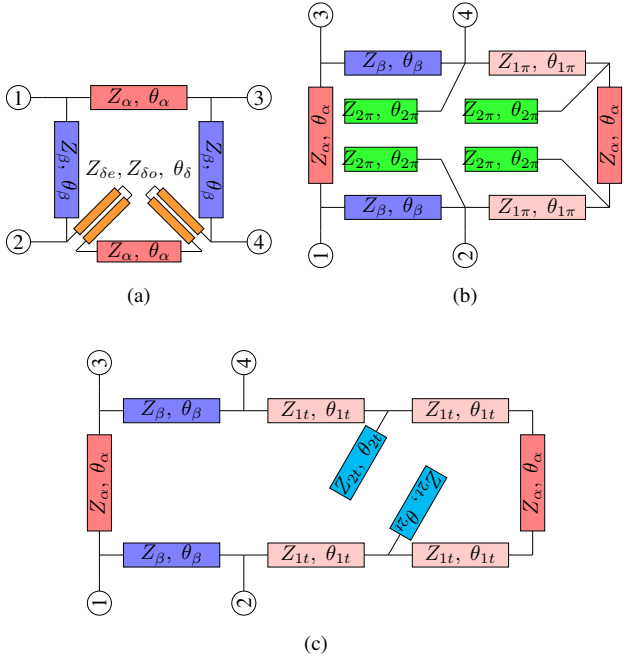
\section{Dual-Band Rat-race Coupler}\label{sec2}
\subsection{Design Equations of Dual-Band RRC}
 The generalised RRC designed using TLs \cite{Sinha2016} can operate in two bands and provide different amplitude-imbalances if $180^{\circ}$ phase-shift is frequency independent. Instead of a frequency-independent phase-inverter or $180^{\circ}$ phase-shifter, we can use two dual-band $90^{\circ}$ phase-shifters. Using the dual-band phase shifters, three dual-band RRCs are proposed and shown in Fig.\ref{fig:dual_rrc}. We considered dual-band C-sections, Pi-structures, and T-structures for the replacement of the $90^{\circ}$ phase shifters.     
 
Here, our objective is to design an RRC which works at two different frequencies, $f_1$ and $f_2=mf_1$, with two different power division ratios $n_1$ and $n_2=kn_1$. In other words, the scattering parameters of the RRC can be written as 
\begin{small}
\begin{equation}
{S}_{\mathit{RRC}} (f_i)=\frac{e^{-j\varphi_i}}{\sqrt{\left(1+n_i\right)}}\begin{bmatrix}
0         &1&\sqrt{n_i}&0\\
1&0         &0         &-\sqrt{n_i}\\
\sqrt{n_i}&0         &0         &1\\
0         &-\sqrt{n_i}&1&0
\end{bmatrix} 
\label{eq:2}
\end{equation}

\end{small}
where $i\in \{1,2\}$ and $\varphi_i$ is the phase-shift at $f_i$ operating frequency. The phase shift $\varphi_i$ depends upon $\theta_{\alpha}$, $\theta_{\beta}$ and $n_i$.  
 To make the coupler operate at two frequencies, we utilize the following properties of an ideal TL and coupled line:
 \begin{itemize}
 \item The electrical lengths $\theta$ are proportional to the frequency $f$ (i.e., $\theta=\frac{2\pi f}{v_g}l$, where $l$ is its physical length, and $v_g$ is the guided velocity). Therefore, the electrical lengths of the TLs hold the following relationship:   
 \begin{equation*}
\theta_{\alpha}(f_2)=m \theta_{\alpha}(f_1);\;\theta_{\beta}(f_2)=m \theta_{\beta}(f_1).
\end{equation*}
 \item  The  characteristic impedances are invariant of frequency, i.e., 
 \begin{equation*}
Z_{\alpha}(f_2)= Z_{\alpha}(f_1);\;Z_{\beta}(f_2)= Z_{\beta}(f_1).
\end{equation*}
 \end{itemize}
 
 After equating characteristic impedances at $f_1$ and $f_2$, the dual-band conditions are obtained as in \eqref{eq:rrc8a} and \eqref{eq:rrc8b}.
 
 The equations \eqref{eq:rrc8a}-\eqref{eq:rrc8b} have to be solved numerically for given values of $m$ and $k$. Once $\theta_{\alpha}$ and $\theta_{\beta}$ are determined using the numerical solution of \eqref{eq:rrc8a}-\eqref{eq:rrc8b}, the characteristic impedances $Z_{\alpha}$ and $Z_{\beta}$ can be calculated using \eqref{eq:rrc_phi1}, \eqref{eq:rrc_phi2}, \eqref{eq:rrc_Za}, and \eqref{eq:rrc_Zb}. 
 
\renewcommand{\thefigure}{S\arabic{figure}}
\setcounter{figure}{0}
\renewcommand{\theequation}{S\arabic{equation}} 
\setcounter{equation}{0}
\renewcommand{\figurename}{Sidebar}
%
\begin{figure}[!h]
\centering
\caption{Design Equations of TL-based dual-band RRC and GPD}
\begin{textblock*}{9cm}(0cm,0.1cm)
\begin{small} 
\begin{itemize}
\item Electrical lengths $\theta_{\alpha}$ and $\theta_{\beta}$ should solved using the following equations
\begin{subequations}
\begin{align}
\frac{\sin m\theta_{\beta}}{\sin \theta_{\beta}}&=\sqrt{k}\frac{\sin m\theta_{\alpha}}{\sin \theta_{\alpha}}\label{eq:rrc8a}\\
\frac{\cos m(\theta_{\alpha}-\theta_{\beta})}{\cos (\theta_{\alpha}-\theta_{\beta})}&= \frac{\cos m(\theta_{\alpha}+\theta_{\beta})}{\cos (\theta_{\alpha}+\theta_{\beta})}.\label{eq:rrc8b}
\end{align}
\label{eq:rrc8}
\end{subequations}
\item The characteristic impedances are obtained using the following equations
\begin{subequations}
\allowbreak
\begin{align}
\cos \varphi_1=&\frac{\sqrt{n_1}\cos\theta_{\alpha}+\cos\theta_{\beta}}{\sqrt{n_1+1}}\label{eq:rrc_phi1}\\
\cos \varphi_2=&\frac{\sqrt{n_2}\cos m\theta_{\alpha}+\cos m\theta_{\beta}}{\sqrt{n_2+1}}\label{eq:rrc_phi2}\\
Z_{\alpha}=&Z_0\frac{\sqrt{1+n_1}}{\sqrt{n_1}}\frac{\sin \varphi_1}{\sin \theta_\alpha}=Z_0\frac{\sqrt{1+n_2}}{\sqrt{n_2}}\frac{\sin \varphi_2}{\sin m\theta_\alpha}\label{eq:rrc_Za}\\
Z_{\beta}=&Z_0\frac{\sqrt{1+n_1}}{1}\frac{\sin \varphi_1}{\sin \theta_\beta}=Z_0\frac{\sqrt{1+n_2}}{1}\frac{\sin \varphi_2}{\sin m\theta_\beta}.\label{eq:rrc_Zb}
\end{align}
\label{eq:rrc1}
\end{subequations}
\end{itemize}
\end{small}
\end{textblock*}
\vspace*{8 cm}
\end{figure}

 The dual-band $90^{\circ}$ phase shifters with desired topological configurations (i.e., C-section, $\Pi$-structure, and T-structure) can be designed using \eqref{eq:C_section}, \eqref{eq:Pi_section}, and \eqref{eq:T_section}. The choices of dual-band topology depend upon the frequency ratio, physical size, and bandwidth. Designers can estimate performance measures via design parameters and circuit analysis results obtained using our proposed design tool \cite{sinha_2024_RRC_GPD_v1.2}.  
 
The proposed dual-band RRC design steps are summarized as follows:

\begin{enumerate}
\allowdisplaybreaks
\item Choose the desired resonance frequencies $f_1$ and $f_2$, power division ratios $n_1$ and $n_2$, and system impedance $Z_0$.
\item Calculate dual-band frequency ratio $m=f_2/f_1$ and power division ratio $k=n_2/n_1$.
\item Solve \eqref{eq:rrc8a} and \eqref{eq:rrc8b} for electrical lengths $\theta_{\alpha}$ and $\theta_{\beta}$  at $f_1$.
\item Calculate the characteristic impedances $Z_{\alpha}$ and $Z_{\beta}$ using \eqref{eq:rrc_Za}-\eqref{eq:rrc_Zb}.
\item Select a dual-band quadrature line (a) C-section, (b) $\Pi$-structure, or (c) T-structure and find the design parameters using \eqref{eq:C_section}, \eqref{eq:Pi_section}, or \eqref{eq:T_section} with $Z_{\gamma}=Z_{\alpha}$. 
\end{enumerate}
\begin{figure}[!h]
\caption{Dual-band Quadrature Phase Shifter}
\centering
\begin{textblock*}{9cm}(0cm,0.2cm)
\begin{small}
\begin{itemize}
\item Design equations of dual-band $90^{\circ}$ phase shifter with $Z_\gamma$ impedance using C-section coupled line can be given as 
\begin{subequations}
\begin{align}
\theta_{\delta}&=\frac{\pi}{m+1}.\label{eq:rrc8c}\\
Z_{\delta e}=&Z_{\gamma}\tan\theta_{\delta} \label{eq:rrc4a}\\
Z_{\delta o}=&Z_{\gamma}\cot\theta_{\delta} ,\label{eq:rrc4b}
\end{align}
\vspace*{-1em}
\label{eq:C_section}
\end{subequations}
\item $Z_{\delta e}$ and $Z_{\delta o}$ are even and odd mode impedances and $\theta_{\delta}$ is the electrical length of the coupled line. 

\item A dual-band $\Pi$-structure equivalent to $90^{\circ}$ phase shifter with impedance $Z_{\gamma}$ can be designed using
\begin{subequations}
\begin{align}
\theta_{1\pi}=\theta_{2\pi}=\frac{\pi}{m+1}\\
Z_{1\pi}=\frac{Z_\gamma}{\sin\theta_{1\pi}}\\
Z_{2\pi}=\frac{Z_{\gamma}\tan\theta_{2\pi}}{\cos\theta_{1\pi}}.
\end{align}
\label{eq:Pi_section}
\end{subequations}
\item $[Z_{1\pi},\theta_{1\pi}]$ and $[Z_{2\pi},\theta_{2\pi}]$ are mainline and open stub of $\Pi$-structure. 
\item  Design equations of dual-band T-type $90^{\circ}$ phase shifter matched to $Z_{\gamma}$ can be written as
\begin{subequations}
\begin{align}
\theta_{1t}=\frac{\theta_{2t}}{2}=\frac{\pi}{m+1}\\
Z_{1t}=\frac{Z_\gamma}{\tan\theta_{1t}}\\
Z_{2t}=\frac{Z_{\gamma}\cos^2\theta_{1t}\tan\theta_{2t}}{\cos2\theta_{1t}},
\end{align}
\label{eq:T_section}
\end{subequations} 
\item $[Z_{1t},\theta_{1t}]$ and $[Z_{2t},\theta_{2t}]$ are mainline and open stub of T-structure. 
\end{itemize}
\end{small}
\end{textblock*}
\vspace*{11.5 cm}
\end{figure}
\renewcommand{\thefigure}{\arabic{figure}}
\setcounter{figure}{1}
\renewcommand{\theequation}{\arabic{equation}} 
\setcounter{equation}{1}
\renewcommand{\figurename}{Fig.}

\begin{figure}
\centering
\includegraphics[width=8cm]{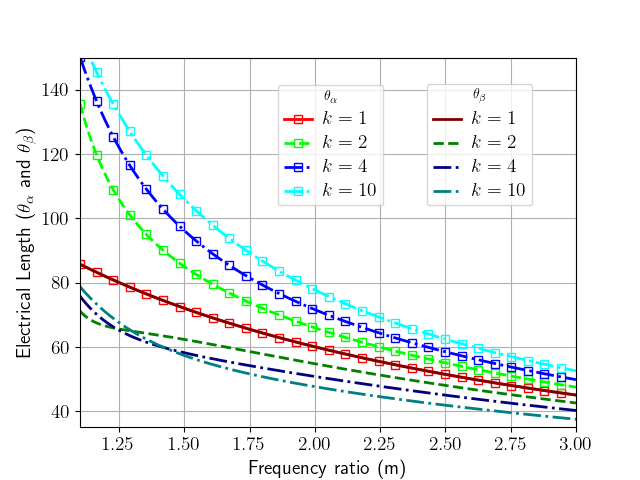}\label{fig:rrc5a}
\caption{Variation of electrical length $\theta_{\alpha}$ (with marker) and $\theta_{\beta}$ (without marker) with respect to frequency ratio $m$ for different values of $k=n_2/n_1$. }
\label{fig:el_lenght}
\end{figure}
\begin{figure*}[!t]
\vspace*{-1em}
\centering
\subfloat[]{\includegraphics[width=6cm]{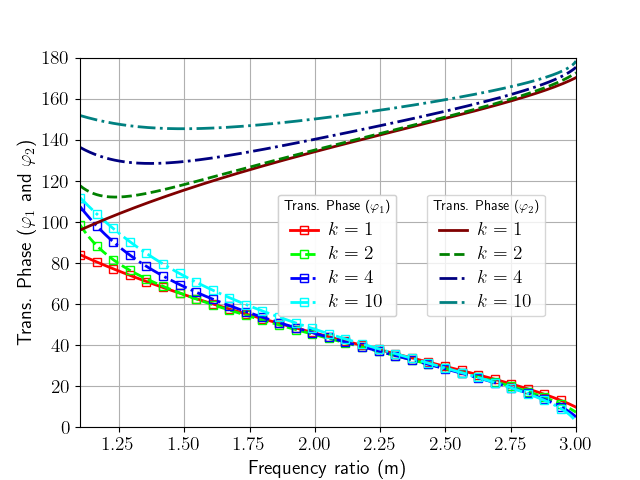}\label{fig:rrc_phia}}
\subfloat[]{\includegraphics[width=6cm]{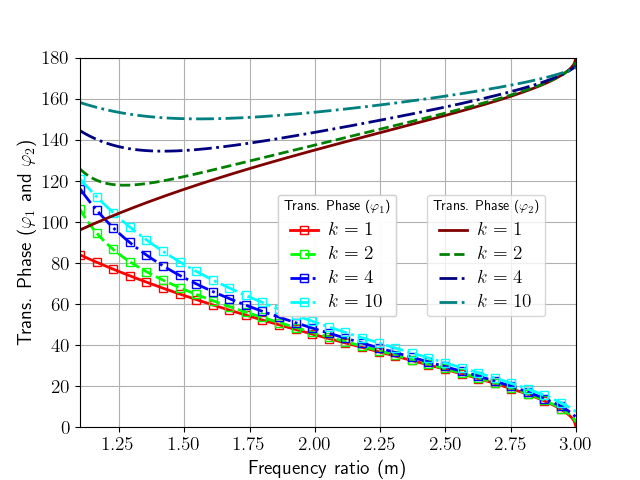}\label{fig:rrc_phib}}
\subfloat[]{\includegraphics[width=6cm]{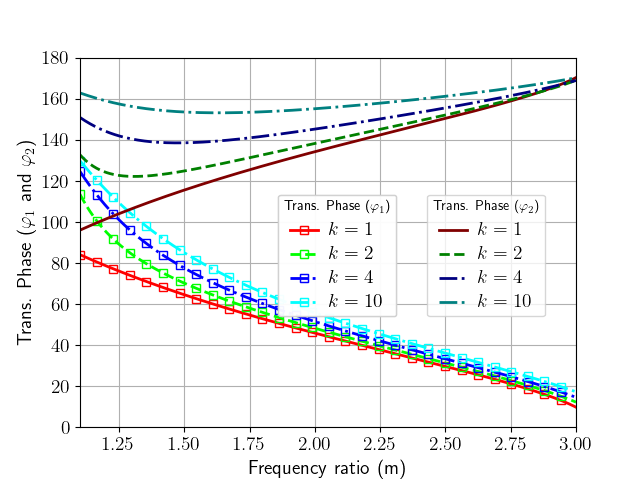}\label{fig:rrc_phic}}
\caption{Variation of phase shift $\varphi_1$ (with marker) and $\varphi_2$ (without marker) with respect to frequency ratio $m$ for different values of $k=n_2/n_1$ and (a) $n_1=1/2$,(b) $n_1=1$, (c) $n_1=2$. }
\label{fig:rrc_phi}
\end{figure*}
\begin{figure*}[!t]
\vspace*{-1em}
\centering
\subfloat[]{\includegraphics[width=6cm]{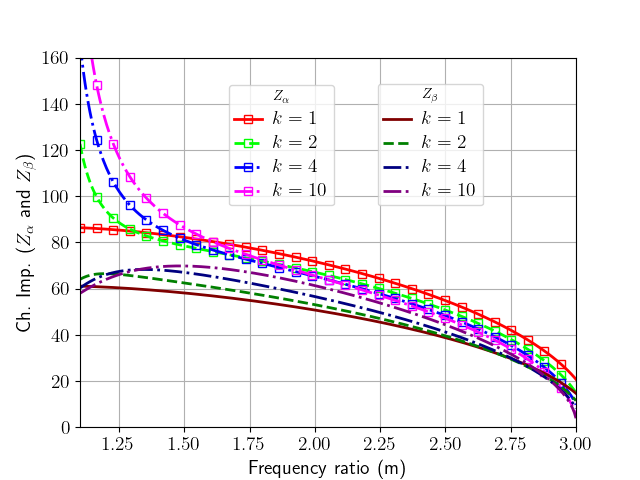}\label{fig:rrc_ZaZb_a}}
\subfloat[]{\includegraphics[width=6cm]{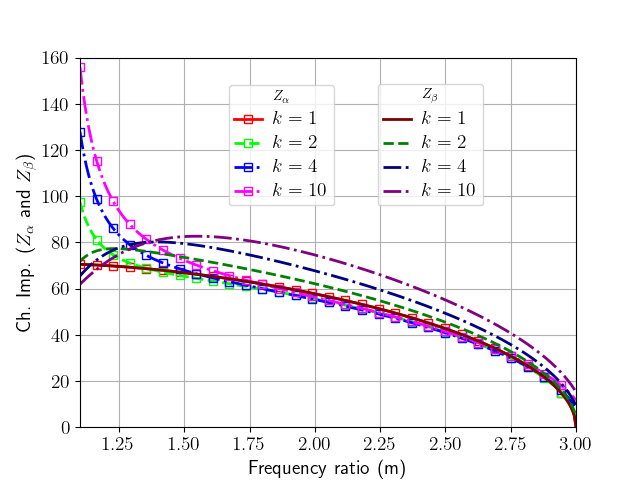}\label{fig:rrc_ZaZb_b}}
\subfloat[]{\includegraphics[width=6cm]{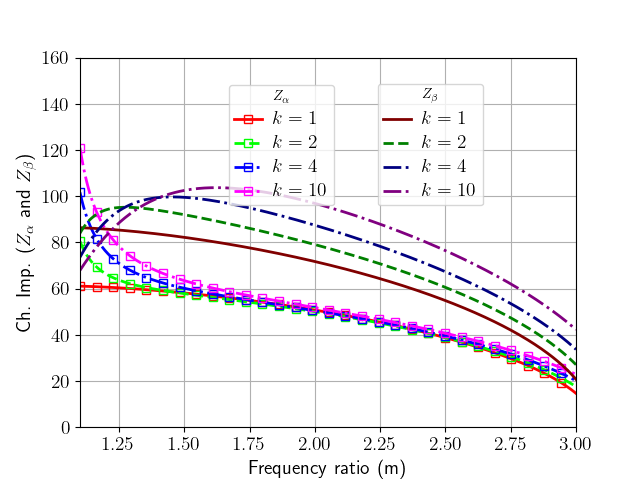}\label{fig:rrc_ZaZb_c}}
\caption{Variation of characteristic impedance $Z_{\alpha}$ (with marker) and $Z_{\beta}$ (without marker) with respect to frequency ratio $m$ for different values of $k=n_2/n_1$ and (a) $n_1=1/2$,(b) $n_1=1$, (c) $n_1=2$. }
\label{fig:rrc_ZaZb}
\end{figure*}
\begin{figure}
\centering
\includegraphics[width=7cm]{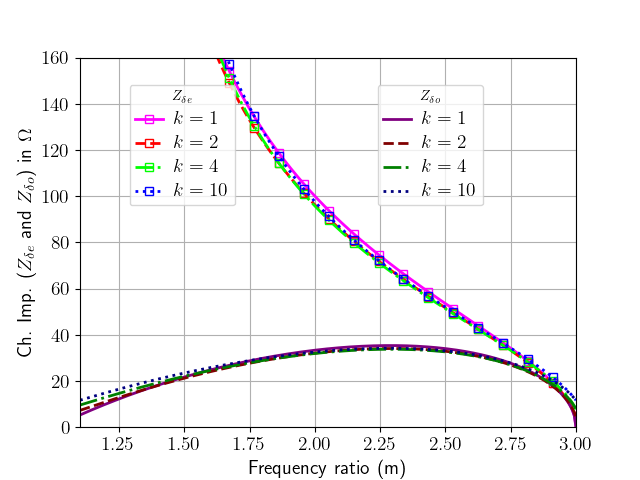}
\caption{Variation of characteristic impedance $Z_{\delta e}$ (with marker) and $Z_{\delta o}$ (without marker) with respect to frequency ratio $m$ for different values of $k=n_2/n_1$ and  $n_1=1$. }
\label{fig:rrc_ZeZo_b}
\end{figure}

\begin{figure}
\centering
\includegraphics[width=7cm]{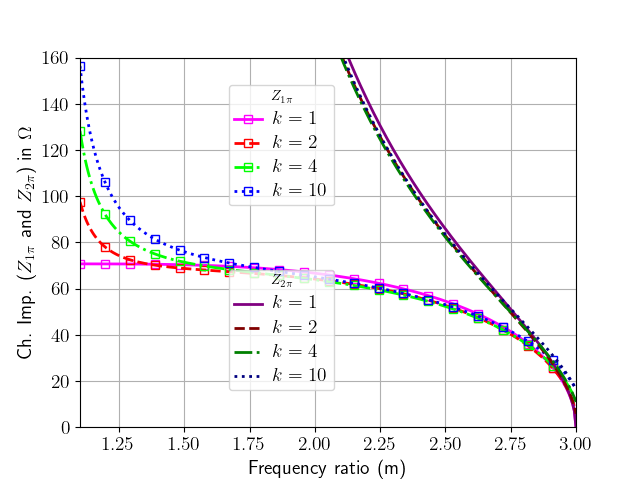}
\caption{Variation of characteristic impedance $Z_{1\pi}$ (with marker) and $Z_{2\pi}$ (without marker) with respect to frequency ratio $m$ for different values of $k=n_2/n_1$ and $n_1=1$. }
\label{fig:rrc_Zpi_b}
\end{figure}

\begin{figure}[!h]
\vspace*{-1em}
\centering
\includegraphics[width=7cm]{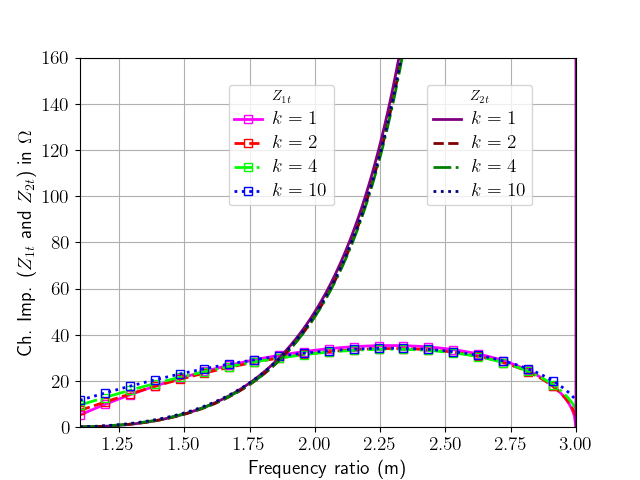}
\caption{Variation of characteristic impedance $Z_{1t}$ (with marker) and $Z_{2t}$ (without marker) with respect to frequency ratio $m$ for different values of $k=n_2/n_1$ and $n_1=1$. }
\label{fig:rrc_Zt_b}
\end{figure}
%
%
\subsection{Design Parameters of Dual-Band Design}\label{sec2c}
In this subsection, we discuss the design parameters of the proposed dual-band couplers (Fig. \ref{fig:dual_rrc}). From the dual-band design equations \eqref{eq:rrc8a} and \eqref{eq:rrc8b}, it is clear that electrical lengths $\theta_{\alpha}$ and $\theta_{\beta}$ are functions of frequency ratio $m =f_2/f_1$ and the power division ratio $k=n_2/n_1$. In order to demonstrate the variation of the electrical parameters, a continuous variation of $1.1\leq m \leq 3.0$ and four discrete values of $k=[1,2,4,10]$ are chosen. The dual-band design equations \eqref{eq:rrc8a} and \eqref{eq:rrc8b} are solved using the Python SciPy module, and the results are plotted in Fig. \ref{fig:el_lenght}. The results shown in Fig. \ref{fig:el_lenght} are equally applicable for $k= [1, 1/2, 1/4, 1/10]$; however, the values of $\theta_{\alpha}$ and $\theta_{\beta}$ are interchangeable with $\theta_{\alpha} (1/k)= \theta_{\beta} (k)$ and $\theta_{\beta} (1/k)= \theta_{\alpha} (k)$ (i.e., $\theta_{\alpha} (k=1/2)= \theta_{\beta} (k=2)$ and  $\theta_{\beta} (k=1/2)= \theta_{\alpha} (k=2)$). Fig. \ref{fig:el_lenght} shows $\theta_{\alpha}=\theta_{\beta}=\frac{\pi}{1+m}$ when $k=1$. This result can also be concluded from \eqref{eq:rrc8a} and \eqref{eq:rrc8b} with $k=1$.  The electrical lengths of the C-section, $\Pi$ structure, and T-structure are functions of frequency ratio $m$ only and are independent of $k$.  Note that the electrical lengths $\theta_{\alpha}$ and $\theta_{\beta}$  shown in Fig. \ref{fig:el_lenght} are calculated at the first resonance frequency $f_1$. 

Once the dual-band electrical lengths in \eqref{eq:rrc8a} and \eqref{eq:rrc8b}  are solved, we can obtain the phase shifts $\varphi_1$ and $\varphi_2$ using \eqref{eq:rrc_phi1} and \eqref{eq:rrc_phi2}. The variations of $\varphi_1$ and $\varphi_2$ with respect to $m$ are shown in Fig.\ref{fig:rrc_phi} for different values of $k=[1,2,4,10]$ and $n_1=[1/2,1,2]$. We can observe that the phase shifts are not proportional to the electrical lengths of the TLs of the RRC.   

After determining the electrical lengths $\theta_{\alpha}$ and $\theta_{\beta}$, and phase shifts $\varphi_1$ and $\varphi_2$, the characteristic impedances $Z_{\alpha}$ and $Z_{\beta}$ are determined using \eqref{eq:rrc1}. The characteristic impedances $Z_{\alpha}$ and $Z_{\beta}$ are functions of $n_1$, $k$, and $m$. In order to understand the variation of $Z_{\alpha}$ and $Z_{\beta}$ with respect to $m$, three different values of $n_1=[1/2,1,2]$ are chosen. Considering $Z_0=50\;\Omega$, the transmission line impedances $Z_{\alpha}$ and $Z_{\beta}$ are plotted in Figs. \ref{fig:rrc_ZaZb_a}, \ref{fig:rrc_ZaZb_b},   and \ref{fig:rrc_ZaZb_c} for $n_1=1/2$, $n=1$, and $n=2$, respectively. Figs. \ref{fig:rrc_ZaZb_a}, \ref{fig:rrc_ZaZb_b},   and \ref{fig:rrc_ZaZb_c}  are equally applicable to the case of $k=[1, 1/2, 1/4, 1/10]$; however, the value of $\theta_{\alpha}$ and $\theta_{\beta}$ will interchange with $\theta_{\alpha} (1/n_1,1/k)= \theta_{\beta} (n_1,k)$ and $\theta_{\beta} (1/n_1, 1/k)= \theta_{\alpha} (n_1, k)$. For example Fig. \ref{fig:rrc_ZaZb_a} also represents the case where $n_1=2$, where $\theta_{\alpha}$ (without marker) and $\theta_{\beta}$ (with marker) varies with respect to $m$ for $k=[1, 1/2, 1/4, 1/10]$.

The even and odd mode characteristic impedances of the C section coupled line $Z_{\gamma e}$ and $Z_{\gamma o}$ can be calculated using \eqref{eq:C_section} with $Z_{\gamma}=Z_{\alpha}$. The coupled line impedances $Z_{\gamma e}$ and $Z_{\gamma o}$ are plotted in Fig. \ref{fig:rrc_ZeZo_b} for $n_1=1$. By observing the impedance values, we can conclude that a dual-band C-section is suitable for the frequency ratio $2\leq m \leq 2.75$.  

The mainline and shunt line impedances $Z_{1\pi}$ and $Z_{2\pi}$ of the $\Pi$-structure are calculated using \eqref{eq:Pi_section}. The variations of $Z_{1\pi}$ and $Z_{2\pi}$ are shown in Fig. \ref{fig:rrc_Zpi_b}. Fig. \ref{fig:rrc_Zpi_b} shows that a dual-band $\Pi$-structure is suitable for $2.25\leq m \leq 2.9$. 

We can calculate mainline and shunt line impedances $Z_{1t}$ and $Z_{2t}$ of the T-structure using \eqref{eq:T_section}. Fig. \ref{fig:rrc_Zt_b} shows the variations of $Z_{1t}$ and $Z_{2t}$. Thus, a dual-band T-structure is suitable for $1.75\leq m \leq 2.25$. 
   
 With the discussed design parameters, we can design a dual-band RRC with different power divisions. In the following section, we will show that RRC design equations are equally applicable to a GPD.  
 
\begin{figure}

\vspace*{-2em}
\centering
\subfloat[]{
\begin{tikzpicture}
\node[rotate=0,scale=0.6] at (0,0){
\begin{tikzpicture}[scale=1]
\tikzmath{\x1 = 6*0.6;\s=0.6; } 
\TL{(0,0)}{0}{\s}{red!50}{$Z_{\alpha},\; \theta_{\alpha}$};
\port{(0,0)}{180}{\s}{1}{0.5};
\port{(\x1,0)}{0}{\s}{3}{0.5};
\TL{(0,0)}{-90}{\s}{blue!50}{$Z_{\beta},\; \theta_{\beta}$};
\TL{(\x1,0)}{-90}{\s}{blue!50}{$Z_{\beta},\; \theta_{\beta}$};
\port{(0,-\x1)}{180}{\s}{2}{0.5};
\draw (\x1,-\x1)--(\x1+0.5,-\x1);
\draw (\x1+0.5,-\x1) to[R,l=$Z_0$](\x1+0.5,-\x1-1.3);
\draw (\x1+0.5,-\x1-1.3) -- (\x1+0.5,-\x1-1.3) node[ground]{}; 
\TL{(0,-1*\x1)}{0}{\s}{red!50}{$Z_{\alpha},\; 180^{\circ}+\theta_{\alpha}$};
\end{tikzpicture}};
\end{tikzpicture}
\label{fig:pd1a}}
\subfloat[]{
\begin{tikzpicture}
\node[rotate=0,scale=0.6] at (-1,0){
\begin{tikzpicture}[scale=1]
\tikzmath{\x1 = 3;\s=0.5;\x2=1.5; } 
\TL{(0,\x2)}{0}{\s}{red!50}{$Z_{\alpha},\; \theta_{\alpha}$};
\port{(0,0)}{180}{\s}{1}{0.5};
\TL{(0,-\x2)}{0}{\s}{blue!50}{$Z_{\beta},\; \theta_{\beta}$};
\port{(\x1,-\x2)}{-90}{\s}{2}{0.5};
\port{(\x1,\x2)}{90}{\s}{3}{0.5};
\TL{(\x1,-\x2)}{0}{\s}{red!50}{$Z_{\alpha},\; \theta_{\alpha}$};
\TL{(\x1,\x2)}{0}{\s}{blue!50}{$Z_{\beta},\; \theta_{\beta}$};
\TL{(2*\x1,\x2)}{-90}{\s}{green!50}{$Z_{\gamma},\; 180^{\circ}$};
\draw (0,-\x2)--(0,\x2);

\draw (2*\x1,-\x2) to[R,l_=$R_{2}$](2*\x1,-\x2-1.5);
\draw (2*\x1,-\x2-1.5) -- (2*\x1,-\x2-1.5) node[ground]{}; 
\draw (2*\x1,\x2) to[R,l=$R_{3}$](2*\x1,\x2+1.5);
\draw (2*\x1,\x2+1.5) -- (2*\x1,\x2+1.8) node[ground,rotate=180]{}; 
\end{tikzpicture}
};
\end{tikzpicture} 
\label{fig:pd1b}}
\caption{(a) Three port power divider after terminating port-4 of an RRC,  (b) equivalent Gysel power divider}
\label{fig:pd1}
\end{figure}
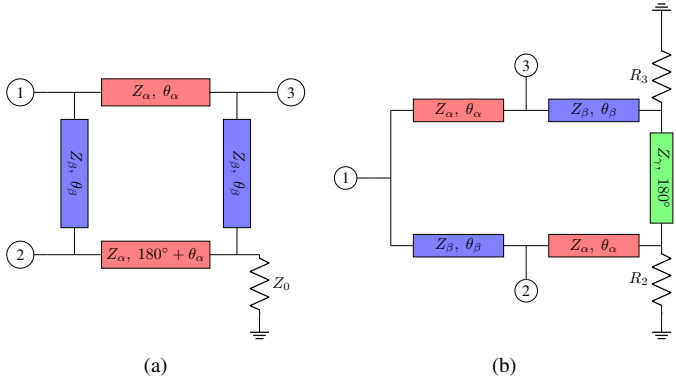
 \section{Rat-Race Coupler to Gysel Power Divider}\label{sec3}
Any design for a four-port RRC is equally applicable to a three-port power divider with output port isolation ($S_{23}=0$). If the isolation port (port-4) of the RRC  is terminated by port impedance $Z_0$, then the circuit acts like a three-port power divider shown in Fig. \ref{fig:pd1a} with scattering matrix 
\begin{equation}
{S}_{{PD}}=\frac{{e}^{-{j\varphi_i}}}{\sqrt{1+n_i}}\begin{bmatrix}
0         &1&\sqrt{n_i}\\
1 &0         &0        \\
\sqrt{n_i} &0         &0         
\end{bmatrix}.\label{eq:pd1}
\end{equation}
where $i\in\{1,2\}$. The design equations of the power divider in Fig. \ref{fig:pd1a} are the same as in \eqref{eq:rrc8} and \eqref{eq:rrc1}. 

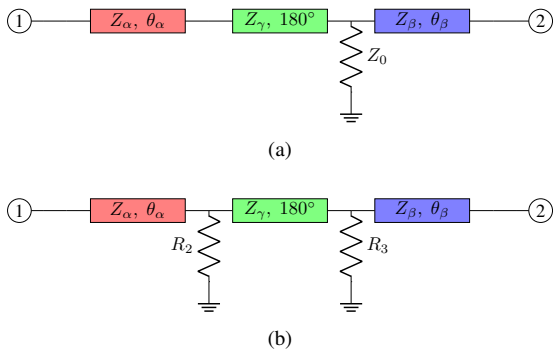
\begin{figure}[!h]
\centering
\subfloat[]{
\begin{tikzpicture}
\node[rotate=0,scale=0.7] at (0,0){
\begin{tikzpicture}[scale=0.9]
\tikzmath{\x1 = 6*0.5;\s=0.5; } 
\TL{(0,0)}{0}{\s}{red!50}{$Z_{\alpha},\; \theta_{\alpha}$};
\port{(0,0)}{180}{\s}{1}{0.5};
\port{(3*\x1,0)}{0}{\s}{2}{0.5};
\TL{(\x1,0)}{0}{\s}{green!50}{$Z_{\gamma},\; 180^{\circ}$};
\TL{(2*\x1,0)}{0}{\s}{blue!50}{$Z_{\beta},\; \theta_{\beta}$};
\draw (2*\x1,0) to[R,l=$Z_0$](2*\x1,-1.5);
\draw (2*\x1,-1.5) -- (2*\x1,-1.5) node[ground]{};
\end{tikzpicture}
};
\end{tikzpicture}
\label{fig:pd2a}}\\
\subfloat[]{
\begin{tikzpicture}
\node[rotate=0,scale=0.7] at (0,0){
\begin{tikzpicture}[scale=0.9]
\tikzmath{\x1 = 6*0.5;\s=0.5; } 
\TL{(0,0)}{0}{\s}{red!50}{$Z_{\alpha},\; \theta_{\alpha}$};
\port{(0,0)}{180}{\s}{1}{0.5};
\port{(3*\x1,0)}{0}{\s}{2}{0.5};
\TL{(\x1,0)}{0}{\s}{green!50}{$Z_{\gamma},\; 180^{\circ}$};
\TL{(2*\x1,0)}{0}{\s}{blue!50}{$Z_{\beta},\; \theta_{\beta}$};
\draw (2*\x1,0) to[R,l=$R_3$](2*\x1,-1.5);
\draw (2*\x1,-1.5) -- (2*\x1,-1.5) node[ground]{};
\draw (\x1,0) to[R,l_=$R_2$](\x1,-1.5);
\draw (\x1,-1.5) -- (\x1,-1.5) node[ground]{};
\end{tikzpicture}
};
\end{tikzpicture}
\label{fig:pd2b}}
\caption{(a) Isolation circuit of the power divider  (b) Modified isolation circuit with two shunt resistances}
\label{fig:pd2}
\end{figure}
\renewcommand{\thefigure}{S\arabic{figure}}
\setcounter{figure}{2}
\renewcommand{\theequation}{S\arabic{equation}} 
\setcounter{equation}{5}
\renewcommand{\figurename}{Sidebar}
\begin{figure}[!h]
\caption{Isolation Resistance of GPD}
\begin{small}
\begin{textblock*}{9cm}(0cm,0.2cm)
\begin{itemize}
\item Isolation circuits of RRC and GPD are related via the following equations
\begin{equation}
\frac{1}{R_2}+\frac{1}{R_3}=\frac{1}{Z_0}.\label{eq:pd3}
\end{equation}
\item One may choose the shunt resistances $R_2$ and $R_3$ as a function of the power division ratio ($n$) such that 
\begin{subequations}
\begin{equation}
R_2=(n+1)Z_0 \label{eq:pd4a}
\end{equation}
\begin{equation}
R_3=\frac{n+1}{n}Z_0. \label{eq:pd4b}
\end{equation}
\end{subequations} 
\end{itemize}
\end{textblock*}
\vspace*{4 cm}
\end{small}
\end{figure}
\renewcommand{\thefigure}{\arabic{figure}}
\setcounter{figure}{9}
\renewcommand{\theequation}{\arabic{equation}} 
\setcounter{equation}{2}
\renewcommand{\figurename}{Fig.}

\begin{figure}[!h]
\centering
\subfloat[]{\begin{tikzpicture}
  \node[rotate=0,scale=0.8] at (0,0) {
\begin{tikzpicture}[scale=0.8]
\tikzmath{\x1 = 3;\s=0.5;\x2=1.0; } 
\TL{(0,\x2)}{0}{\s}{red!50}{$Z_{\alpha},\; \theta_{\alpha}$};
\port{(0,0)}{180}{\s}{1}{0.5};
\TL{(0,-\x2)}{0}{\s}{blue!50}{$Z_{\beta},\; \theta_{\beta}$};
\port{(\x1,-\x2)}{-90}{\s}{2}{0.5};
\port{(\x1,\x2)}{90}{\s}{3}{0.5};
\TL{(\x1,-\x2)}{0}{\s}{red!50}{$Z_{\alpha},\; \theta_{\alpha}$};
\TL{(\x1,\x2)}{0}{\s}{blue!50}{$Z_{\beta},\; \theta_{\beta}$};
\coupledTLr{(2*\x1,0)}{0}{1/3}{green!50}{};
\coupledTLl{(2*\x1,0.3)}{0}{1/3}{green!50}{};
\draw (0,-\x2)--(0,\x2);
\draw (2*\x1,-\x2)--(2*\x1,-0.3);
\draw (2*\x1,\x2)--(2*\x1,0.3);
\draw (2*\x1,-\x2) to[R,l_=$R_{2}$](2*\x1+1.5,-\x2);
\draw (2*\x1+1.5,-\x2) -- (2*\x1+1.8,-\x2) node[ground]{}; 
\draw (2*\x1,\x2) to[R,l=$R_{3}$](2*\x1+1.5,\x2);
\draw (2*\x1+1.5,\x2) -- (2*\x1+1.8,\x2) node[ground]{}; 
\draw[] (2*\x1-1,-0.4) node[] {$Z_{\delta e}, Z_{\delta o},\; \theta_{\delta}$};
\end{tikzpicture}

};
\end{tikzpicture}\label{fig:PD_c}}\\

\subfloat[]{
\begin{tikzpicture}
  \node[rotate=0,scale=0.8] at (0,0) {
\begin{tikzpicture}[scale=0.8]
\tikzmath{\x1 = 3;\s=0.5;\x2=1.0; } 
\TL{(0,\x2)}{0}{\s}{red!50}{$Z_{\alpha},\; \theta_{\alpha}$};
\port{(0,0)}{180}{\s}{1}{0.5};
\TL{(0,-\x2)}{0}{\s}{blue!50}{$Z_{\beta},\; \theta_{\beta}$};
\port{(\x1,-\x2)}{-90}{\s}{2}{0.5};
\port{(\x1,\x2)}{90}{\s}{3}{0.5};
\TL{(\x1,-\x2)}{0}{\s}{red!50}{$Z_{\alpha},\; \theta_{\alpha}$};
\TL{(\x1,\x2)}{0}{\s}{blue!50}{$Z_{\beta},\; \theta_{\beta}$};
\draw (0,-\x2)--(0,\x2);
\draw (3*\x1,-\x2)--(3*\x1,\x2);
\openTL{(3*\x1,0)}{180}{0.7}{green!80}{$Z_{2\pi}/2,\; \theta_{2\pi}$};
\draw (2*\x1,-\x2) to[R,l_=$R_{2}$](2*\x1,-\x2-1.5);
\draw (2*\x1,-\x2-1.5) node[ground]{}; 
\draw (2*\x1,\x2) to[R,l=$R_{3}$](2*\x1,\x2+1.5);
\draw (2*\x1,\x2+1.5) node[ground,rotate=180]{}; 
\draw (2*\x1+1,-\x2-1) to [short] (2*\x1,-\x2);
\draw (2*\x1+1,\x2+1) to [short] (2*\x1,\x2);
\openStub{(2*\x1+1,\x2+1)}{0}{\s}{green!80}{$Z_{2\pi},\; \theta_{2\pi}$};
\openStub{(2*\x1+1,-\x2-1)}{0}{\s}{green!80}{$Z_{2\pi},\; \theta_{2\pi}$};
\TL{(2*\x1,-\x2)}{0}{\s}{pink!80}{$Z_{1\pi},\; \theta_{1\pi}$};
\TL{(2*\x1,\x2)}{0}{\s}{pink!80}{$Z_{1\pi},\; \theta_{1\pi}$};
\end{tikzpicture}

};
\end{tikzpicture}\label{fig:PD_Pi}
}\\

\subfloat[]{
\begin{tikzpicture}
  \node[rotate=0,scale=0.8] at (0,0) {
\begin{tikzpicture}[scale=0.8]
\tikzmath{\x1 = 3;\s=0.5;\x2=1.5; } 
\TL{(0,\x2)}{0}{\s}{red!50}{$Z_{\alpha},\; \theta_{\alpha}$};
\port{(0,0)}{180}{\s}{1}{0.5};
\TL{(0,-\x2)}{0}{\s}{blue!50}{$Z_{\beta},\; \theta_{\beta}$};
\port{(\x1,-\x2)}{-90}{\s}{2}{0.5};
\port{(\x1,\x2)}{90}{\s}{3}{0.5};
\TL{(\x1,-\x2)}{0}{\s}{red!50}{$Z_{\alpha},\; \theta_{\alpha}$};
\TL{(\x1,\x2)}{0}{\s}{blue!50}{$Z_{\beta},\; \theta_{\beta}$};
\TL{(3*\x1,\x2)}{-90}{\s}{pink!80}{$Z_{1t},\; 2\theta_{1t}$};
\draw (0,-\x2)--(0,\x2);
\draw (2*\x1,-\x2) to[R,l_=$R_{2}$](2*\x1,-\x2-1.5);
\draw (2*\x1,-\x2-1.5) node[ground]{}; 
\draw (2*\x1,\x2) to[R,l=$R_{3}$](2*\x1,\x2+1.5);
\draw (2*\x1,\x2+1.5) node[ground,rotate=180]{};
\TL{(2*\x1,-\x2)}{0}{\s}{pink!80}{$Z_{1t},\; \theta_{1t}$};
\TL{(2*\x1,\x2)}{0}{\s}{pink!80}{$Z_{1t},\; \theta_{1t}$};
\draw (3*\x1-0.7,-\x2+1) to [short] (3*\x1,-\x2);
\draw (3*\x1-0.7,\x2-1) to [short] (3*\x1,\x2);
\openTL{(3*\x1-0.7,\x2-1)}{180}{\s}{green!80}{$Z_{2t},\; \theta_{2t}$};
\openTL{(3*\x1-0.7,-\x2+1)}{180}{\s}{green!80}{$Z_{2t},\; \theta_{2t}$};
\end{tikzpicture}

};
\end{tikzpicture}\label{fig:PD_T}
}

\caption{Dual-band Gysel power divider using dual-band $180^{\circ}$ phase-shifters designed using (a) C-sections, (b) $\Pi$-structures, and (c) T-structures}
\label{fig:pd4}
\end{figure}
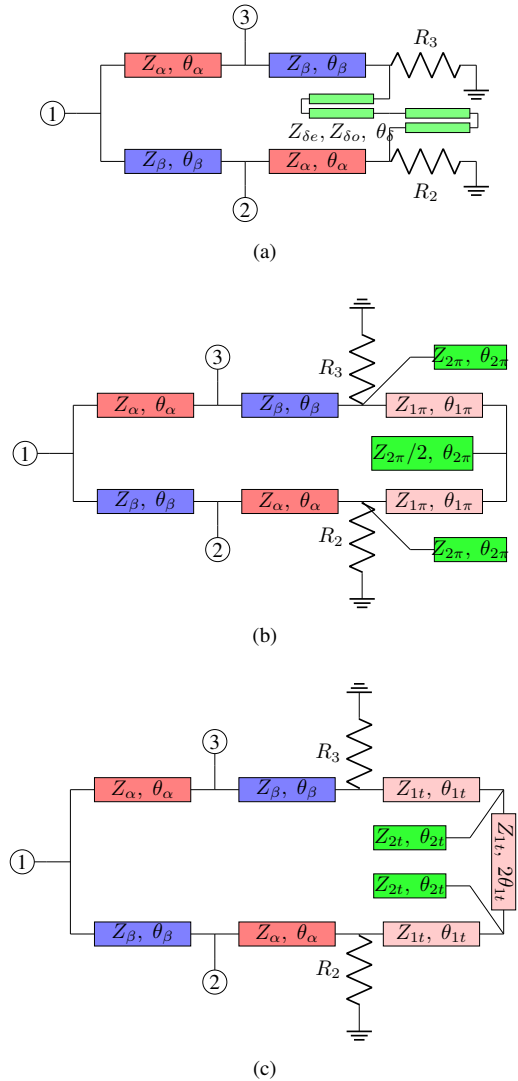

The isolation circuit of the power divider (Fig. \ref{fig:pd1a}) can be represented as the two-port network (TPN) shown in Fig. \ref{fig:pd2a}, where the $\lbrace Z_{\alpha},180^{\circ}+\theta_{\alpha}\rbrace$ line is presented as a cascade connection of the $\lbrace Z_{\alpha},\theta_{\alpha}\rbrace$ and $\lbrace Z_{\gamma},180^{\circ}\rbrace$ line. Here $Z_{\gamma}$ can have any value independent of the port impedance ($Z_0$) or power division ratio ($n$). The isolation circuit shown in  \ref{fig:pd2a} can be further modified into a Gysel power divider like isolation circuit as shown in Fig.  \ref{fig:pd2b}. The TPN shown in Fig.  \ref{fig:pd2b} is equivalent to the TPN shown in Fig.  \ref{fig:pd2a}, if \eqref{eq:pd3} holds.

After replacing the isolation circuit of Fig.  \ref{fig:pd1a} with  \ref{fig:pd2b}, we can design the Gysel-type power divider as shown in Fig.  \ref{fig:pd1b}.  

The $180^{\circ}$ line of the isolation circuit (Fig. \ref{fig:pd1b}) can be implemented using a cascade connection of two identical C-section coupled lines. Henceforth, each C-section coupled line is equivalent to the $\lbrace Z_{\gamma}, 90^{\circ} \rbrace$ line, similar to the case of an RRC with design equation \eqref{eq:C_section}. We can also replace the  $180^{\circ}$ line of a GPD using dual-$\Pi$-structures and dual-T-structures.    

The circuit diagram of the power divider shown in Fig. \ref{fig:pd1b} can be redrawn as the Gysel power divider shown in Fig. \ref{fig:pd4} by replacing the $180^{\circ}$ with (a) C-sections, (b) $\Pi$-structures, and (c) T-structures. The design schematic of Fig. \ref{fig:PD_c} is similar to \cite{Tang2016} with $\theta_{\alpha}=\theta_{\beta}=\frac{\pi}{R+1}$, where the same power division ratio was considered at two bands (i.e., $n_1=n_2$ or $k=1$).The design in Fig. \ref{fig:PD_Pi} has been proposed in \cite{ChenGPD2019}. 

 The design procedure of the proposed dual-band GPD  is summarized as follows:
\begin{enumerate}
\item Choose the desired resonance frequencies $f_1$ and $f_2$,  power division ratios $n_1$ and $n_2$,  system impedance $Z_0$, and impedance of $180^{\circ}$ TL $Z_{\gamma}$.
\item Calculate dual-band frequency ratio $m=f_2/f_1$ and power division ratio $k=n_2/n_1$.
\item Solve \eqref{eq:rrc8a} and \eqref{eq:rrc8b} for electrical lengths $\theta_{\alpha}$ and $\theta_{\beta}$.
\item Calculate the characteristic impedances $Z_{\alpha}$ and $Z_{\beta}$ using \eqref{eq:rrc_Za}-\eqref{eq:rrc_Zb}.
\item Select a dual-band quadrature line (a) C-section, (b) $\Pi$-structure, or (c) T-structure and get the design parameters using \eqref{eq:C_section}, \eqref{eq:Pi_section}, or \eqref{eq:T_section} with desired $Z_{\gamma}$.
 \item Finally, the $R_2$ and $R_3$ of the GPD should be chosen such that \eqref{eq:pd3} holds.
\end{enumerate}
With the above procedure, a dual-band GPD with two different power divisions at two bands can be designed, and the design parameter variations are the same as discussed in  Section-\ref{sec2c}. In addition, the impedances of C-sections, $\Pi$-structures, and T-structures can be scaled for the desired $Z_{\gamma}$. 
The application of two different power-division ratios at two bands is demonstrated in the following Section-\ref{sec4}. 

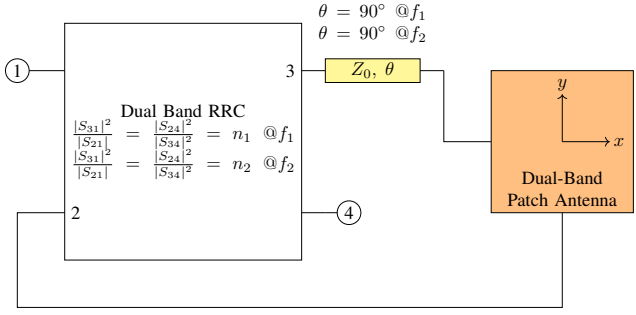
\begin{figure}
\begin{tikzpicture}
\node[rotate=0,scale=0.7] at (0,0) {
\begin{tikzpicture}[rotate=0,scale=0.9]
\tikzmath{\x1 = 2.5;\s=0.5; } 
\draw [] (-\x1,-\x1) rectangle (\x1,\x1) node[pos=0.5,text width=6cm,align=center] {Dual Band RRC\\$\frac{|S_{31}|^2}{|S_{21}|}=\frac{|S_{24}|^2}{|S_{34}|^2}=n_1\;@ f_1$\\$\frac{|S_{31}|^2}{|S_{21}|}=\frac{|S_{24}|^2}{|S_{34}|^2}=n_2\;@ f_2$};
\TL{(\x1,\x1-1)}{0}{\s}{yellow!50}{$Z_{0},\; \theta$};
\draw[] (\x1+1.5,\x1) node[text width=3cm,align=center] {$\theta=90^{\circ}\; @ f_1$\\$\theta=90^{\circ}\; @ f_2$};
\port{(-\x1,\x1-1)}{180}{\s}{1}{0.5};
\port{(\x1,-\x1+1)}{0}{\s}{4}{0.5};
\draw[] (-\x1,-\x1+1) node[right] {2};
\draw[] (\x1,\x1-1) node[left] {3};
\draw [] (-\x1,-\x1+1)--(-\x1-1,-\x1+1)--(-\x1-1,-\x1-1)--(\x1+5.5,-\x1-1)--(\x1+5.5,-1.5);
\draw[] (\x1+3,\x1-1)--(\x1+3,0)--(\x1+4,0);
\draw [fill={orange!50}](\x1+4,1.5) rectangle (\x1+4+3,-1.5); 
\draw [](5.5+\x1,-1) node[text width=3cm,align=center]{Dual-Band \\Patch Antenna};
\draw [->](5.5+\x1,0) --(5.5+\x1+1,0) node[pos=1.2]{$x$};
\draw [->](5.5+\x1,0) --(5.5+\x1,1) node [pos=1.2]{$y$};
\end{tikzpicture}
};
\end{tikzpicture}
\caption{Schematic of a dual-band circular polarized antenna using an RRC with different axial ratios at two bands.}
\label{fig:cp1}
\end{figure} 

\begin{figure}[!h]
\begin{tikzpicture}
\node[rotate=0,scale=0.8] at (0,0) {
\begin{tikzpicture}[rotate=0,scale=0.8]
\tikzmath{\x1 = 2.0;\s=0.5; } 
\draw [] (-\x1,-\x1) rectangle (\x1,\x1) node[pos=0.5,text width=6cm,align=center] {Dual Band PD\\$\frac{|S_{31}|^2}{|S_{21}|}=n_1\;@ f_1$\\$\frac{|S_{31}|^2}{|S_{21}|}=n_2\;@ f_2$};
\TL{(\x1,\x1-1)}{0}{\s}{yellow!50}{$Z_{0},\; \theta$};
\draw[] (\x1+1.5,\x1) node[text width=3cm,align=center] {$\theta=90^{\circ}\; @ f_1$\\$\theta=90^{\circ}\; @ f_2$};
\port{(-\x1,0)}{180}{\s}{1}{0.5};
\draw[] (\x1,-\x1+1) node[left] {2};
\draw[] (\x1,\x1-1) node[left] {3};
\draw [] (\x1,-\x1+1)--(\x1+1,-\x1+1)--(\x1+1,-\x1)--(\x1+5.5,-\x1)--(\x1+5.5,-1.5);
\draw[] (\x1+3,\x1-1)--(\x1+3,0)--(\x1+4,0);
\draw [fill={orange!50}](\x1+4,1.5) rectangle (\x1+4+3,-1.5); 
\draw [](5.5+\x1,-1) node[text width=3cm,align=center]{Dual-Band \\Patch Antenna};
\draw [->](5.5+\x1,0) --(5.5+\x1+1,0) node[pos=1.2]{$x$};
\draw [->](5.5+\x1,0) --(5.5+\x1,1) node [pos=1.2]{$y$};
\end{tikzpicture}
};
\end{tikzpicture}
\caption{Schematic of a dual-band circularly polarized antenna using a GPD with different axial ratios at two bands.}
\label{fig:cp2}
\end{figure}
\begin{figure}
\vspace*{-1.0em}
\centering
\includegraphics[width=8 cm]{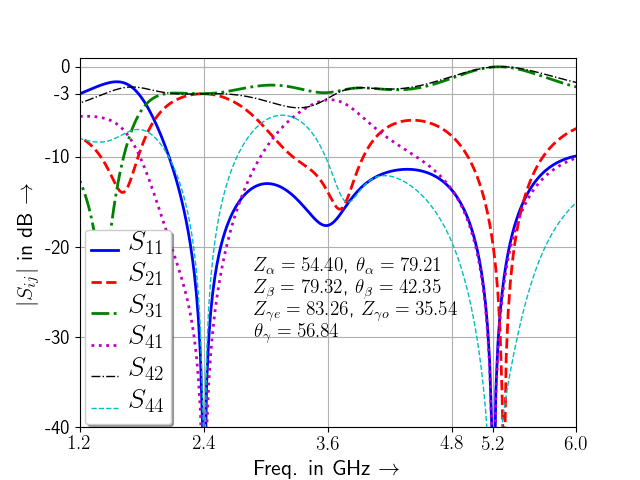}
\caption{{Amplitude response of dual-band RRC with $n_1=0$ dB and $n_2=20$ dB @ $f_1=2.4$ GHz and $f_2=5.2$ GHz}}
\label{fig:Amp_res}
\end{figure}

\begin{figure}
\vspace*{-1em}
\centering
\includegraphics[width=6 cm]{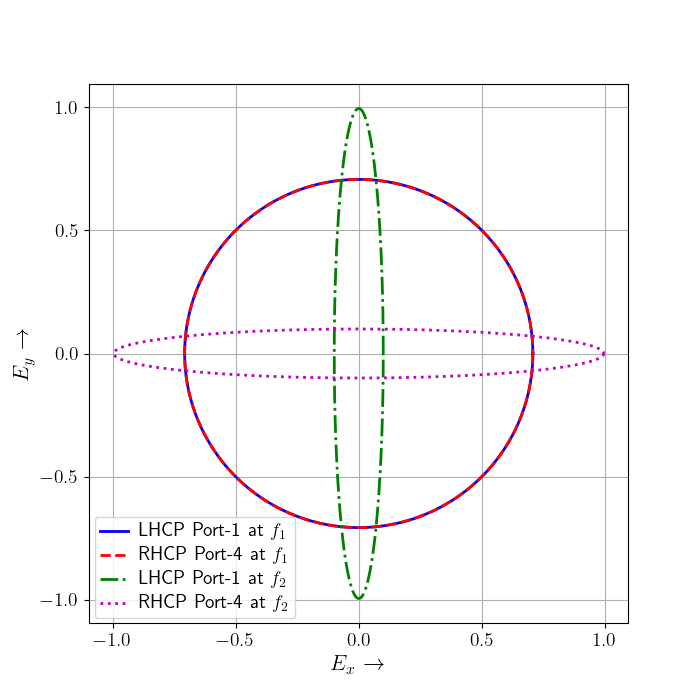}
\caption{{Different polarizations of the CP antenna in  Fig. \ref{fig:cp1} using a dual-band RRC with amplitude response as in Fig. \ref{fig:Amp_res}}}
\label{fig:Pol_div}
\end{figure}
\section{Application of Different Power Division Ratio at Two Band}\label{sec4}
A possible application of dual-band RRC with different power division ratios at two bands is different polarization control at two bands. The schematic of the idea is shown in Fig.\ref{fig:cp1}, where dual-band circular/elliptical polarization (CP) control of a patch antenna is presented. For circular/elliptical polarization, a quadrature-phase difference between the orthogonal excitation signals (i.e., x and y-directed feeds) is required. The dual-band quarter-wave line connected to port-3 provides a quadrature-phase difference between the output signals. The axial ratio of the antenna depends upon the power division ratios at two bands. A GPD-based CP control schematic is shown in Fig. \ref{fig:cp2}.  The polarization can be generated through the system shown in Fig. \ref{fig:cp1}:
\begin{itemize}
\item LHCP (CW) $\vec{E}=\sqrt{n_1}\hat{x}+j\hat{y}$ with axial ratio $\frac{|E_x|^2}{|E_y|^2}=n_1$ at $f_1$ and LHCP (CW) $\vec{E}=\sqrt{n_2}\hat{x}+j\hat{y}$ with axial ratio $\frac{|E_x|^2}{|E_y|^2}=n_2$ at $f_2$ when port-1 is excited. If a quadrature phase difference is not present,  then linear polarization $\vec{E}=\sqrt{n_1}\hat{x}+\hat{y}$ at $f_1$ and linear polarization $\vec{E}=\sqrt{n_2}\hat{x}+\hat{y}$ at $f_2$ when port-1 is excited.  
\item  RHCP (CCW) $\vec{E}=\hat{x}-j\sqrt{n_1}\hat{y}$ with axial ratio $\frac{|E_x|^2}{|E_y|^2}=\frac{1}{n_1} $ at $f_1$ and RHCP (CCW) $\vec{E}=\hat{x}-j\sqrt{n_2}\hat{y}$ with axial ratio $\frac{|E_x|^2}{|E_y|^2}=\frac{1}{n_2}$ at $f_2$ when port-4 is excited. If a quadrature phase difference is not present,  then linear polarization $\vec{E}=\hat{x}-\sqrt{n_1}\hat{y}$ at $f_1$ and linear polarization $\vec{E}=\hat{x}-\sqrt{n_2}\hat{y}$ at $f_2$ when port-4 is excited. 
\end{itemize}

{For example, consider a dual-band RRC with $n_1=0 $ dB and $n_2=20 $ dB implemented with the proposed technique at frequency $f_1=2.4$ GHz and $f_2=5.2$ GHz. The design parameters are calculated as  $[\theta_\alpha, \theta_\beta, Z_{\alpha}, Z_{\beta}, \theta_{\gamma}, Z_{\gamma e}, Z_{\gamma o}]=$ $[79.21^{\circ}, 42.35^{\circ}, 54.4\;\Omega, 79.32\;\Omega, 56.84^{\circ}, 83.26\;\Omega, 35.54\;\Omega]$ at 2.4 GHz. The amplitude response of the RRC is shown in Fig. \ref{fig:Amp_res} . Fig. \ref{fig:Amp_res} shows that the RRC provides 0 dB and 20 dB amplitude imbalance at 2.4 GHz and 5.2 GHz. If we use this RRC in Fig. \ref{fig:cp1}, four different polarizations can be obtained for port-1 and port-4 excitation at  2.4 GHz and 5.2 GHz. At 2.4 GHz, pure LHCP and RHCP can be obtained for port-1 and port-4 excitations, respectively. However, at 5.2 GHz, we can suppress x-polarization by 20 dB compared to y-polarization for the port-1 excitation. On the other hand, y-polarization is suppressed by 20 dB from x-polarization for the port-4 excitation.}

\begin{figure*}[!t]
\centering
\includegraphics[width=13 cm]{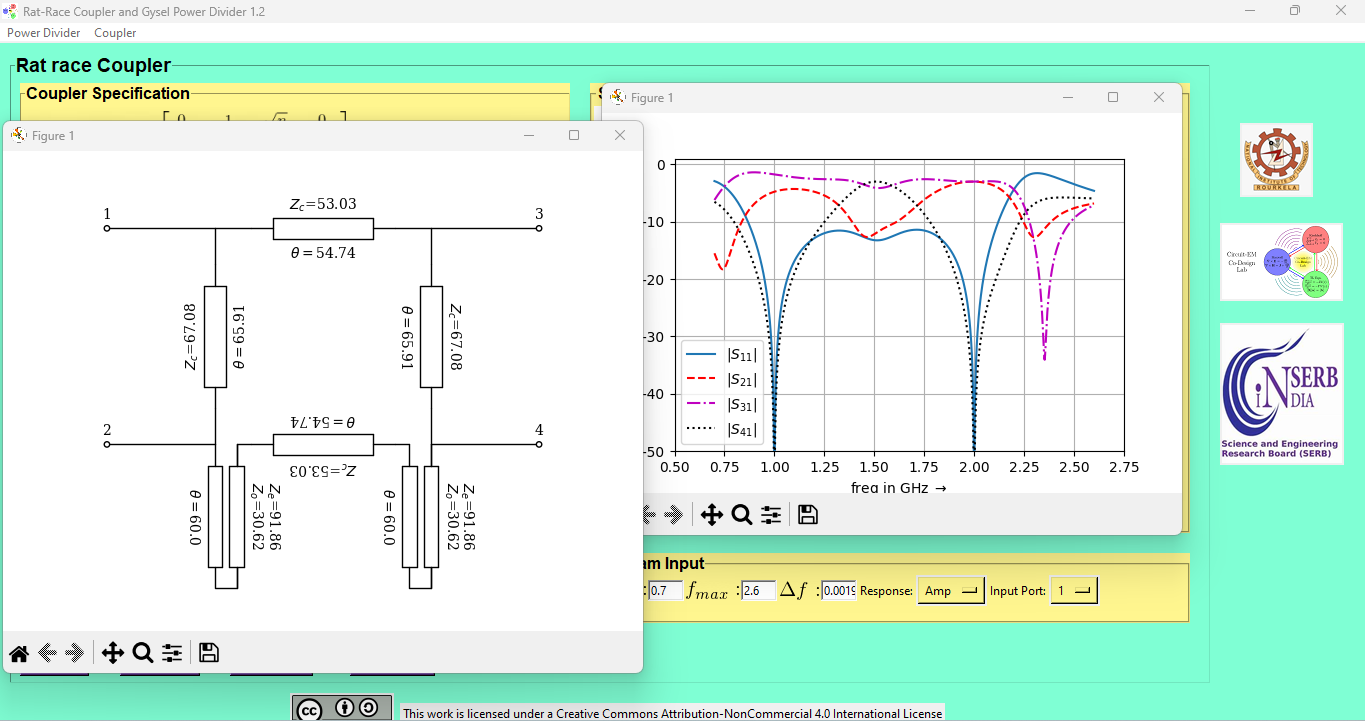}
\caption{Python-based Windows application available at \url{https://zenodo.org/records/11199141} to design and simulate the proposed GPDs and RRCs.}
\label{fig:gui}
\end{figure*}

\begin{figure*}[!t]
\vspace*{-1em}
\centering
\subfloat[]{\label{fig:6a}\includegraphics[width=6cm]{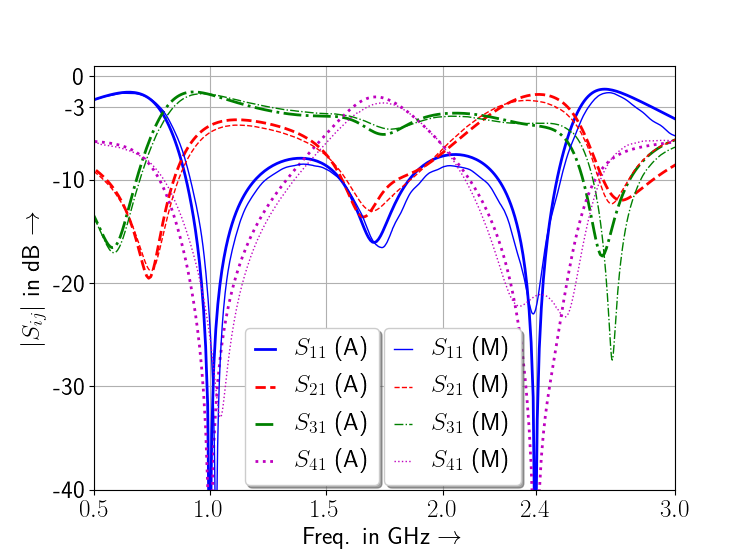}}
\subfloat[]{\label{fig:6b}\includegraphics[width=6cm]{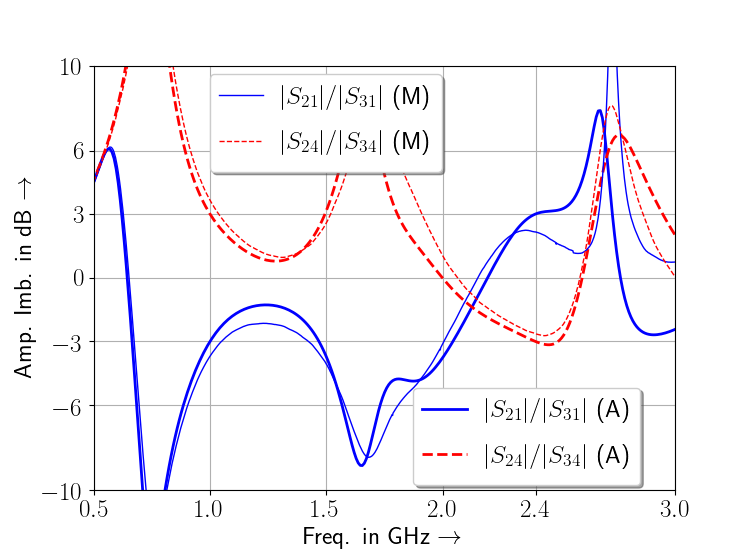}}
\subfloat[]{\label{fig:6c}\includegraphics[width=6cm]{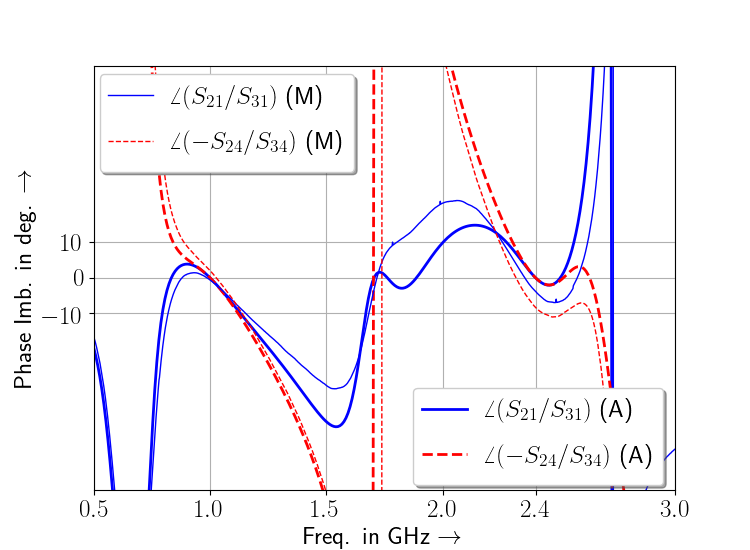}}
\caption{ Predicted (thick line) and measured (thin line) result of the proposed coupler(a) Amplitude response,(b) Amplitude imbalance, (c) Phase imbalance }
\label{fig:6}
\end{figure*}
\section{Python-based Windows Application}\label{sec:cad}
In order to enhance the reproducibility of the design data and analysis shown here, we have developed a Python-based Windows application. The application window is shown in Fig. \ref{fig:gui}. The free software is available at Zenodo \cite{sinha_2024_RRC_GPD_v1.2} under Creative Commons Attribution 4.0 International.  The application is capable of designing RRCs and GPDs for single and dual-band operation. The user can set the design frequency $f_0$ or $f_1$ and $f_2$, power division ratio $n$ or $n_1$ and $n_2$ based on their requirements. There are different choices for $180^{\circ}$ (two $90^{\circ}$) phase shifters available. The phase shifters are C-section, $\Pi$-section, T-section, and TL-section. The user can understand the benefits and costs of using a particular phase shifter by changing the different phase shifters. The software can generate design schematics and analysis results, which can be saved for future uses. Some of the design and analysis results are summarized in the supplementary material.

 {In the following section, we will discuss the practical design of a dual-band RRC using the C-sections. The design of dual-band RRCs and GPDs with different phase shifters can be carried out by interested readers using the proposed CAD application \cite{sinha_2024_RRC_GPD_v1.2}.}   

\begin{figure}
\vspace*{-1em}
\centering
\subfloat[]{\label{fig:7a}\includegraphics[width=3.5cm,height=3.2cm]{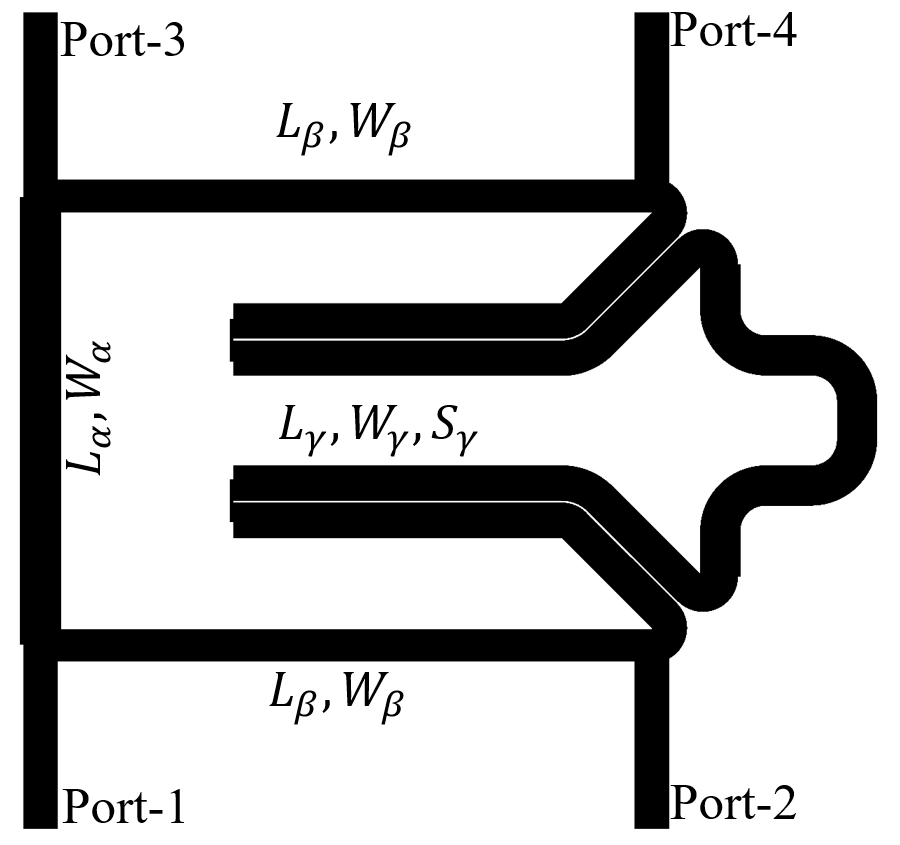}}
\subfloat[]{\label{fig:7b}\includegraphics[width=3.5cm,height=3.2cm]{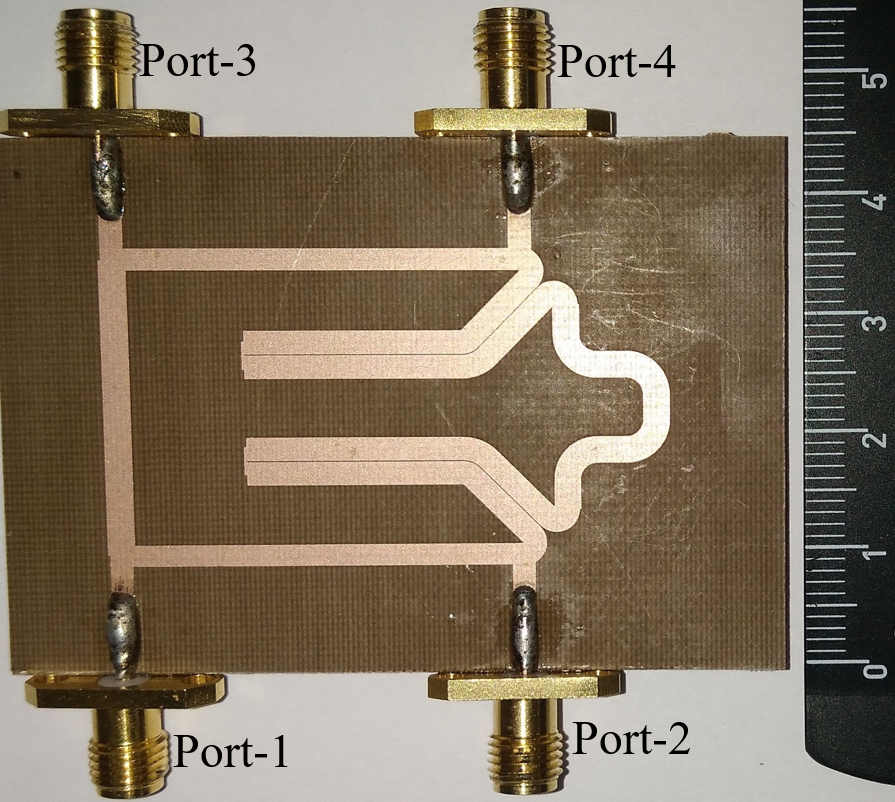}}
\caption{(a)Layout of the proposed RRC; (b) fabricated prototype}
\label{fig:7}
\end{figure} 
\section{Design Example}\label{sec5}
In order to verify the proposed concept, a dual-band rat-race coupler is considered here with resonance frequencies $f_1=1$ GHz and $f_2=2.4$  GHz, and power division ratio $n_1=3$ dB and $n_2=-3$  dB (i.e., $k=-6$ dB). The electrical parameters are calculated using the proposed design equations as: $\theta_{\alpha}(f_1)=46.04^{\circ}$, $\theta_{\beta}(f_1)=60.61^{\circ}$, $\theta_{\delta}(f_1)=52.94^{\circ}$, $Z_{\alpha}=44.80\; \Omega$, $Z_{\beta}=52.34\;\Omega$, $Z_{\delta e}=59.32\; \Omega$ and $Z_{\delta o}=33.83\; \Omega$.

The proposed coupler is implemented using microstrip technology on a Taconic RF 30 substrate of thickness 30 mil with a dielectric constant of 3.0 and loss tangent 0.0014. The physical dimensions of the coupler are given as $L_{\alpha}=24.53$ mm $W_{\alpha}=2.26$ mm $L_{\beta}=32.58$ mm $W_{\beta}=1.78$ mm $L_{\gamma}=25.73$ mm $W_{\gamma}=1.94$ mm  $S_{\gamma}=0.10$ mm. The layout of the proposed coupler is shown in Fig. \ref{fig:7a}, and the fabricated prototype is shown in Fig. \ref{fig:7b}. The coupler occupies an area of $46.97 \;\text{mm} \times 26.45\;\text{mm}$. The simulated (ideal TL model) and measured (using Anritsu MS46522B network analyzer) results are shown in Fig. \ref{fig:6}. The amplitude response of the proposed coupler shown in Fig. \ref{fig:6a}, shows variations of S-parameters $S_{11}$, $S_{21}$, $S_{31}$, and $S_{14}$ with respect to frequency. The amplitude imbalance or amplitude ratio of the coupler is shown in Fig. \ref{fig:6b}. Fig. \ref{fig:6c} shows the phase imbalance or phase difference between the output ports. {The measured amplitude imbalances are $|S_{21}|-|S_{31}|=-3.69$ dB and $|S_{24}|-|S_{34}|=3.65$ dB at 1 GHz, and $|S_{21}|-|S_{31}|=2.16$ dB and $|S_{24}|-|S_{34}|=-2.65$ dB at 2.4 GHz.} The measured phase imbalances are $\angle S_{21}- \angle S_{31}=-0.65^{\circ}$ and $\angle S_{24}- \angle S_{34}=-177.97^{\circ}$ at 1 GHz, and $\angle S_{21}- \angle S_{31}=-3.9^{\circ}$ and $\angle S_{24}- \angle S_{34}=-188.44^{\circ}$ at 2.4 GHz. The measured results are in good agreement with the predicted ones around the first band of frequency. However, the measured results deviate from the simulated ones around the second band of frequency. This is due to the frequency-dependent behavior of the even-odd mode impedances of the coupled line. Other causes of the differences may be fabrication defects and conductors, dielectric, connectors, and radiation losses.   

\section{Conclusion}\label{sec6}
{A general design procedure for a dual-band rat-race coupler with different power dividing ratios is presented in this article. The electrical properties of the coupler are utilized for dual-band operation. The rat-race coupler can be turned into a power divider after modification of the isolation circuit. The design equations of the coupler and power divider are identical. The concept of dual-band polarization control of a dual-band patch antenna is introduced by utilizing a dual-band coupler or power divider with different power dividing ratios. A user-friendly application for circuit design and analysis is also provided. A dual-band rat-race coupler prototype is developed, and measured results are compared with the simulated results.}
\color{black}

\vspace*{-1em}
\section*{Acknowledgement}
This work was partially supported by the Science and Engineering Research Board, India, under Grant SRG/2022/000156. The author would like to thank Prof. Ick-Jae Yoon of Electrical Engineering, Chungnam National University, for providing support in the fabrication and measurement of the dual-band RRC prototype. The author would also like to thank the reviewers and editors for their valuable suggestions for improving this
article. This article has supplementary downloadable
material available at \url{https://zenodo.org/records/11230820}, provided by the author.
\color{black}

\bibliographystyle{MyIEEEtran}
\bibliography{ref_1}

\end{document}